\documentclass[twocolumn]{aastex63}
\usepackage{lineno}

\usepackage{graphicx}	
\usepackage{amsmath}	
\usepackage{amssymb}	
\usepackage{CJK}

\submitjournal{ApJ}
\shorttitle{Lensed FRBs to be detected by BURSTT}
\shortauthors{Ho et al.}

\graphicspath{{./}{figures/}}

\begin{document}
\begin{CJK*}{UTF8}{gbsn}
\title{Future Constraints on Dark Matter with Gravitationally Lensed Fast Radio Bursts Detected by BURSTT}

\correspondingauthor{Simon C.-C. Ho}
\email{Simon.Ho@anu.edu.au}
\author[0000-0002-8560-3497]{Simon C.-C. Ho(何建璋)}
\affiliation{Research School of Astronomy and Astrophysics, The Australian National University, Canberra, ACT 2611, Australia}
\affiliation{Institute of Astronomy, National Tsing Hua University, 101, Section 2. Kuang-Fu Road, Hsinchu, 30013, Taiwan}
\author[0000-0001-7228-1428]{Tetsuya Hashimoto}
\affiliation{Department of Physics, National Chung Hsing University, 145 Xingda Rd., South Dist., Taichung 40227, Taiwan}

\author[0000-0002-6821-8669]{Tomotsugu Goto}
\affiliation{Institute of Astronomy, National Tsing Hua University, 101, Section 2. Kuang-Fu Road, Hsinchu, 30013, Taiwan}

\author[0000-0002-1110-7378]{Yu-Wei Lin}
\affiliation{Department of Physics, National Tsing Hua University, 101, Section 2. Kuang-Fu Road, Hsinchu, 30013, Taiwan}

\author[0000-0001-9970-8145]{Seong Jin Kim}
\affiliation{Institute of Astronomy, National Tsing Hua University, 101, Section 2. Kuang-Fu Road, Hsinchu, 30013, Taiwan}

\author[0000-0003-2792-4978]{Yuri Uno}
\affiliation{Department of Physics, National Chung Hsing University, 145 Xingda Rd., South Dist., Taichung 40227, Taiwan}

\author[0000-0003-4512-8705]{Tiger Y.-Y. Hsiao}
\affiliation{Department of Physics and Astronomy, Johns Hopkins University, Baltimore, MD 21218}

\begin{abstract}

Understanding dark matter is one of the most urgent questions in modern physics. A very interesting candidate is primordial black holes \citep[PBHs;][]{Carr2016}. For the mass ranges of $< 10^{-16} M_{\odot}$ and $> 100 M_{\odot}$, PBHs have been ruled out. However, they are still poorly constrained in the mass ranges of $10^{-16} - 100 M_{\odot}$ \citep{Belotsky2019}. Fast radio bursts (FRBs) are millisecond flashes of radio light of unknown origin mostly from outside the Milky Way. Due to their short timescales, gravitationally lensed FRBs, which are yet to be detected, have been proposed as a useful probe for constraining the presence of PBHs in the mass window of $<$ $100M_{\odot}$ \citep{Munoz2016}. Up to now, the most successful project in finding FRBs has been CHIME. Due to its large field of view (FoV), CHIME is detecting at least 600 FRBs since 2018. However, none of them is confirmed to be gravitationally lensed \citep{Leung2022}. Taiwan plans to build a new telescope, BURSTT dedicated to detecting FRBs. Its survey area will be 25 times greater than CHIME. BURSTT can localize all of these FRBs through very-long-baseline interferometry (VLBI). We estimate the probability to find gravitationally lensed FRBs, based on the scaled redshift distribution from the latest CHIME catalog and the lensing probability function from \citet{Munoz2016}. BURSTT-2048 can detect $\sim$ 24 lensed FRBs out of $\sim$ 1,700 FRBs per annum. With BURSTT's ability to detect nanosecond FRBs, we can constrain PBHs to form a part of dark matter down to $10^{-4}M_{\odot}$.

\end{abstract}



\section{Introduction} \label{sec:intro}
Understanding the nature of dark matter is one of the most interesting questions in modern physics. Broadly, dark matter searches have focused on the subatomic matter (Weakly Interacting Massive Particles or WIMPs) and objects in the astrophysical mass domain (Massive Compact Halo Objects, or MACHOs). In the search for WIMPs, considerable efforts have been made. WIMPs are hypothetical particles that can be considered a candidate for dark matter. Recent sub-atomic experiments such as the LUX-ZEPLIN \citep[LZ; ][]{Aalbers2022} experiment, DarkSide-50 experiment \citep{DarkSide2022}  and XENONnT experiment \citep{XENON2023} carried by particle physicists have placed constraints on the parameter space of WIMPs \citep{Kimball2023}. Nevertheless, no conclusive evidence for WIMPs as a significant contributor to DM has been found to date. A particular method of gravitational microlensing has been used in recent decades to search for dark matter around our Milky Way galaxy in the form of MACHOS \citep{Alcock2000}.
For example, microlensing surveys have derived constraints on mass ranges in the range of 10$^{-7}$ to 10 $M_{\odot}$ \citep[e.g.,][]{Alcock2000, Tisserand2007}. Moreover, constraints from the cosmic microwave background exclude masses $>$100 $M_{\odot}$ \citep{Ali2017}. 
In MACHO and Exp\'erience pour la Recherche d'Objets Sombres (EROS) \citep{Alcock2000, Tisserand2007}, steady background sources are monitored on timescales of days to weeks while their apparent brightness is monitored over time. Primordial Black Hole (PBH) distributions within the Local Group can thus be constrained by these searches. 
On the other hand, gravitational wave observations from mergers of compact binaries \citep{LIGO2016} have recently revived interest in the possibility that dark matter is mainly compact matter \citep{Laha2020}, such as PBHs for which the most promising mass range for dark matter is 10-100 $M_{\odot}$.

One of the methods for detecting dark compact objects is through gravitational lensing of radio transients, such as fast radio bursts (FRBs) \citep{Munoz2016, Eichler2017, Katz2020, Wucknitz2021}. FRBs are mysterious millisecond radio signals mostly emitted by extragalactic sources \citep{Cordes2019}. Although we are not yet able to explain how these millisecond bursts originate or emit, their cosmological distance and abundance make them a good candidate for time-domain gravitational lensing searches. \citep{Munoz2016, Eichler2017, Katz2020, Wucknitz2021}.
As the FRB propagates around a foreground mass, coherent multiple images of the FRB are generated with small differences in the timing, which can be resolved in the time domain as interference fringes \citep{Kader2022}.
Meanwhile, FRBs have been found in other galaxies \citep[e.g.,][]{Chatterjee2017, Ravi2019, Bannister2019}. Using FRBs as probes, one could constrain cosmological abundances of PBHs rather than local abundances in the future.
The current challenge in searching for gravitationally lensed FRBs is that the detection rate of FRBs is still too low. Since the first discovery of FRBs more than a decade ago \citep{Lorimer2007}, astronomers have observed more and more FRBs every year. Among all of the search projects, the most successful project in finding FRBs has been CHIME \citep{CHIME2018}. Due to its large field of view (FoV), CHIME has outperformed other radio arrays detecting about 600 FRBs since 2018 \citep{CHIME2021}, which is about 5/6 of the current observed number of FRBs. However, with this number of FRBs, no gravitationally lensed FRB has been confirmed. \citet{Leung2022, Kader2022} have observed no lenses from 172 bursts of 114 independent events of the CHIME catalog. They have placed an upper limit on the constraint of dark matter made of compact objects, such as PBHs.
Laser Interferometer Gravitational-Wave Observatory (LIGO)/VIRGO interferometer has suggested a hint to the dark matter fraction \citep{LIGO2016}. Based on LIGO's result, \citet{Bird2016} suggest the black hole merger rate is consistent with all dark matter being $\sim$ 30 $M_{\odot}$ black holes. If this is the case, about 1$-$2\% of all FRBs will pass close enough to such a PBH to be microlensed \citep{Munoz2016} and 10$^{4}$ FRBs would yield about 100$-$200 microlensed FRBs. If, on the other hand, no lenses are found in $10^4$ FRBs, the dark matter fraction in PBHs can be constrained to be below 1\% \citep{Munoz2016}.

Taiwan plans to build a new radio telescope, the Bustling Universe Radio Survey Telescope in Taiwan \citep[BURSTT;][]{Lin2022} which is dedicated to detecting FRBs with an accurate localization capability ($\sim$ 1"). BURSTT will be the next frontier telescope in FRB science after CHIME. Because of the telescope's unique fisheye design, it will have a massive 1 steradian(sr) FoV, meaning its survey area will be 25 times greater than CHIME, covering more FRB-like events. BURSTT will have a 256-antenna array in the first phase and it will be extended to a 2048-antenna array in the second phase. This improves the sensitivity by about 8 times. Moreover, BURSTT will conduct long-term monitoring observations to prevent missing any repeating FRBs. Regarding the time resolution, BURSTT has a ring buffer to record the voltage data, which the timing resolution is adjustable to be about a nanosecond (H.-H. Lin 2022, private communication). If we detect any gravitationally lensed nanosecond FRB, we can use it to constrain dark matter down to $10^{-4} M_{\odot}$ of the lensing mass (see Eq. \ref{eq:02} for more detail). The proposed BURSTT opens up the exciting possibility for the searches of gravitationally lensed FRBs which can help us constrain the fraction of dark matter better. In this work, we focus on the future prospect of BURSTT-2048. We aim to give an estimation of how many lensed FRBs will be detected by BURSTT-2048.

This paper is composed as follows; we describe the gravitational lensing model in Section \ref{sec:model}. The predicted number of lensed FRBs to be detected by BURSTT and the calculation are presented in Section \ref{sec:modelling BURSTT}. Discussion in \ref{sec:discussion} and conclusion are in Section \ref{sec:conclusion}. Throughout this paper, we adopt the AB magnitude system and assume a cosmology with H$_0$ = 70 km s$^{-1}$Mpc$^{-1}$, $\Omega_\Lambda$ = 0.7, and $\Omega_\text{M}$ = 0.3  \citep{Spergel2003}.

\section{Gravitational lensing model} \label{sec:model}
In this section, for strong lensing by compact objects, we calculate the optical depth by calculating microlensing effects on a given FRB. These results allow us to calculate the number of lensed FRBs that can be expected when all dark matter is PBHs with a combination of different redshift distributions. We consider compact objects that may be modeled using a point-mass lens, which is the simplest kind of lens model from \citet{Munoz2016}, in which a PBH of mass $M_{L}$. We do not consider a mass spectrum but only fixed mass cases for $M_{L}$, which can be modeled as point lenses around the Einstein radius:

\begin{equation} \label{eq:01}
    \theta_{E}=2 \sqrt{\frac{G M_{L}}{c^{2}} \frac{D_{L S}}{D_{S} D_{L}}} \ .
\end{equation}

The (angular-diameter) distances from the source, to the lens, and between the source and the lens are represented by $D_{S}, D_{L}$, and $D_{L S}$, respectively \citep{Takahashi2003}. With a point lens, two images are formed at positions $\theta_{\pm}=\left(\beta \pm \sqrt{\beta^{2}+4 \theta_{E}^{2}}\right) / 2$, where $\beta$ corresponds to impact angle \citep{Munoz2016}.
The time delay between these two images is

\begin{equation} \label{eq:02}
    \Delta t=\frac{4 G M_{L}}{c^{3}}\left(1+z_{L}\right)\left[\frac{y}{2} \sqrt{y^{2}+4}+\log \left(\frac{\sqrt{y^{2}+4}+y}{\sqrt{y^{2}+4}-y}\right)\right] ,
\end{equation}

where $y \equiv \beta / \theta_{E}$ represents the normalized impact parameter, and the lens redshift is $z_{L}$ \citep{Munoz2016}. In this equation, the time delay, $\Delta t$ is mainly affected by the lensing mass, $M_{L}$. In addition, we follow \citet{Munoz2016} to define the $R_{f}$ as the flux ratio of the magnifications of both images $\mu_{+}$ and $\mu_{-}$; i.e.,

\begin{equation} \label{eq:03}
R_{f} \equiv\left|\frac{\mu_{+}}{\mu_{-}}\right|=\frac{y^{2}+2+y \sqrt{y^{2}+4}}{y^{2}+2-y \sqrt{y^{2}+4}}>1 .
\end{equation}

Following \citet{Munoz2016}, an FRB must meet three conditions to be considered strongly lensed. First, the brighter image between the two pulses has a signal-to-noise ratio of 10 or higher \citep{Petroff2014}. Second, the observed time delay is longer than some reference time $\overline{\Delta t}$ (e.g., observational resolution), and therefore a lower bound will be placed on the impact parameter $y>y_{\min }\left(M_{L}, z_{L}\right)$, as calculated by Eq. \ref{eq:02}. Finally, we require that the flux ratio $R_{f}$ is lower than a critical value for $\bar{R}_{f}$ (which we take to be redshift independent) so that both events are observed. By doing this, the impact parameter is forced to be smaller than \citep{Munoz2016}:
\begin{equation} \label{eq:04}
y_{\max}=\left[\left(1+\bar{R}_{f}\right) / \sqrt{\bar{R}_{f}}-2\right]^{1 / 2}.
\end{equation}

\citet{Munoz2016} calculated the probability for an FRB to be lensed as follows. The lensing optical depth of a source at redshift $z_{S}$ is given by

\begin{equation} \label{eq:05}
\tau\left(M_{L}, z_{S}\right)=\int_{0}^{z_{S}} d \chi\left(z_{L}\right)\left(1+z_{L}\right)^{2} n_{L} \sigma\left(M_{L}, z_{L}\right) .
\end{equation}

 Here, $\chi(z)$ represents the comoving distance at redshift $z$, $n_{L}$ is the comoving number density of lenses comoving, and $\sigma$ is the lensing cross-section of a point lens with mass $M_{L}$, expressed as an annulus between the maximum and minimum impact parameters by the amount of mass. The equation is as follows

\begin{equation} \label{eq:06}
\sigma\left(M_{L}, z_{L}\right)=\frac{4 \pi G M_{L}}{c^{2}} \frac{D_{L} D_{L S}}{D_{S}}\left[y_{\max }^{2}-y_{\min }^{2}\left(M_{L}, z_{L}\right)\right] .
\end{equation}

Using the Hubble parameter both at the redshift of the lens, $H\left(z_{L}\right)$, and present $H_{0}$, we can recast Eq. \ref{eq:05} as

\begin{equation} \label{eq:07}
\begin{aligned}
\tau\left(M_{L}, z_{S}\right)=& \frac{3}{2} f_{\mathrm{DM}} \Omega_{c} \int_{0}^{z_{S}} d z_{L} \frac{H_{0}^{2}}{c H\left(z_{L}\right)} \frac{D_{L} D_{L S}}{D_{S}} \\
& \times\left(1+z_{L}\right)^{2}\left[y_{\max }^{2}-y_{\min }^{2}\left(M_{L}, z_{L}\right)\right],
\end{aligned}
\end{equation}

where $\Omega_{c}=0.24$ is the cold-dark-matter density at present and $\Omega_{m}= \Omega_{c} + \Omega_{b}$. The only remaining dependence on the lens mass $M_{L}$ is through $y_{\min }$.

\begin{figure*}
\centering
	\includegraphics[width=\linewidth]{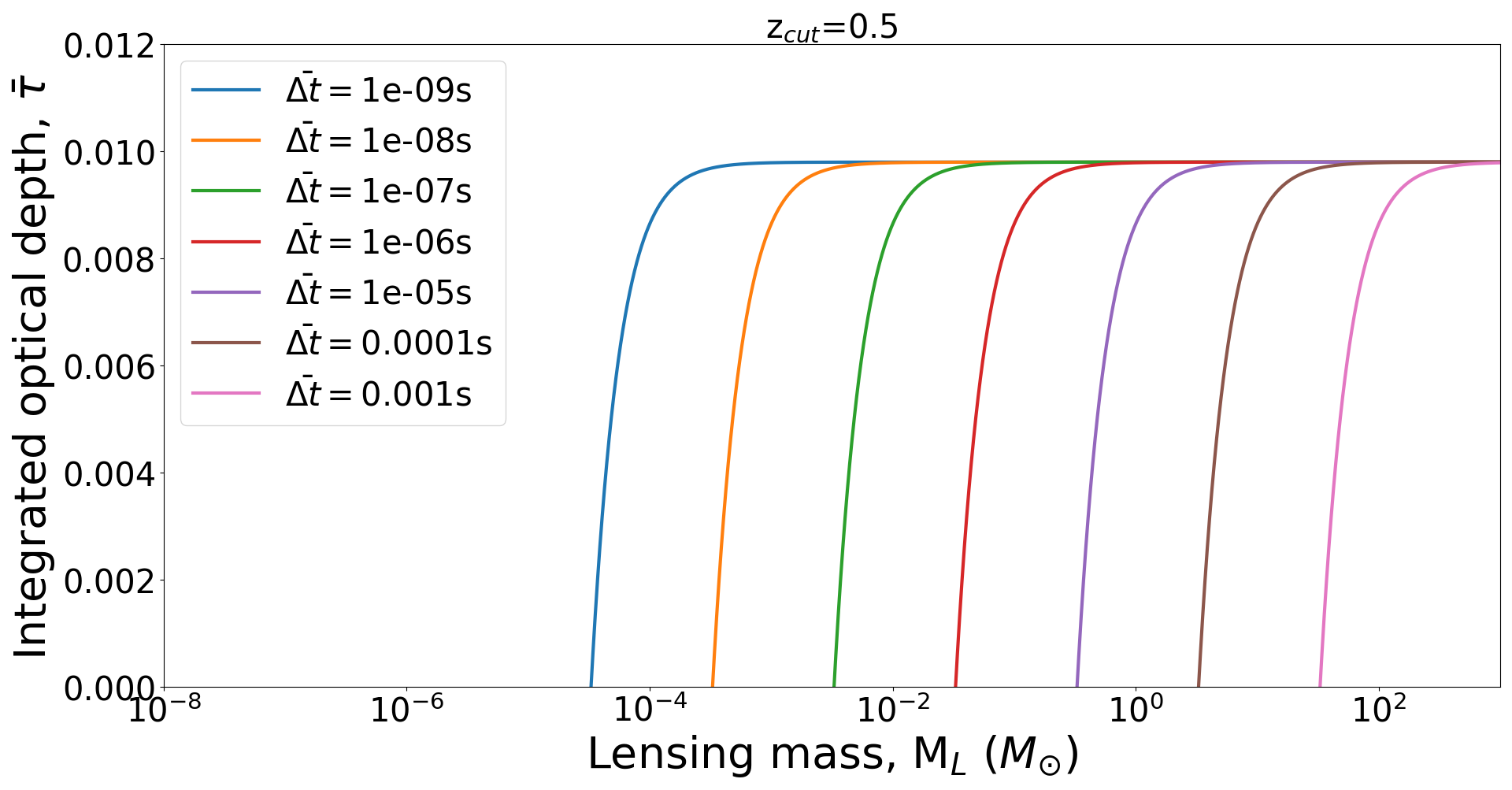}
 
    \caption{Integrated optical depth, with a cutoff at $z_{\text {cut}}=0.5$. We apply the same $z_{\text {cut }}$ as \citet{Munoz2016}.} For the different colors of the solid curve, we require a time delay of $\bar{\Delta t}$ = 1 ms, 0.1 ms, 10 $\mu$s, 1 $\mu$s, 0.1 $\mu$s, 10 ns, and 1 ns, respectively. In all cases, $f_{\mathrm{DM}}=1$. The x-axis is in logarithmic scale.
    \label{fig:tau}
\end{figure*}

Given a normalized observed distribution function $N(z)$ for FRBs  with redshift, $z$, \citep{CHIME2021,Hashimoto2022}, we can calculate their integrated optical depth $\bar{\tau}\left(M_{L}\right)$, due to compact objects of mass $M_{L}$, as
\begin{equation} \label{eq:08}
\begin{aligned}
\bar{\tau}\left(M_{L}\right)=\int d z \, \tau\left(z, M_{L}\right) N(z) .
\end{aligned}
\end{equation}
Here, $\tau$ is included because it becomes approximately the same as the lensing probability when $\tau$ $\ll$ 1.
We show this quantity in Fig. \ref{fig:tau} for different time delays with a redshift cutoff at $z_{\rm cut}=0.5$ and $f_{\mathrm{DM}}$ = 1 (assuming all the dark matter was composed of PBHs). \citet{Munoz2016} adopted a $z$ = 0.5 cutoff redshift because their redshift distribution function fits well with the observed redshift distribution of FRB samples \citep[FRBCAT,][]{Petroff2016} if $z$ = 0.5 is chosen. We applied the redshift cutoff at $z_{\rm cut}=0.5$ as \citet{Munoz2016} since the redshift distributions from both works peak at $z$ $\sim$ 0.5. In this figure, $N(z)=1$ is assumed. We take an empirically derived $N(z)$ from CHIME FRBs into account in Section \ref{sec:lensed FRBs detected by BURSTT}. It is shown that the distribution shifts to a smaller lensing mass as the time delay reduces. It is important to note that, even if all the dark matter consisted of a single mass $M_{L}$, there will be no unique value for the lensing time delays induced on FRBs because of the different redshifts of lenses and impact parameters.


\section{Modelling the FRB events to be detected by BURSTT}\label{sec:modelling BURSTT}

BURSTT is a proposed instrument tailored for detecting FRB-like events with accurate arcsecond localization and high cadence \citep{Lin2022}. BURSTT will observe 25 times more sky than CHIME, because of its unique fisheye design and FoV of 1 sr. BURSTT's modular nature allows it to be expanded by adding additional antennas or outrigger stations. The main station could be upgraded from 256 antennas to 2048 antennas in the next phase. The system equivalent flux density (SEFD) will improve about 8 times from $\sim$ 5,000 Jy in the first phase to $\sim$ 600 Jy in the second phase. This would improve the sensitivity to detect more bursts, which improves the detection rate. At the same time, BURSTT is an interferometer with baselines of over 100 km and has an astrometric accuracy for FRB localizations of better than 1", more precise than that of CHIME \citep[2- and 25-arcsec precision between the CHIME and the CHIME Pathfinder;][]{Leung2021}, ASKAP \citep[arcsecond accuracy, ][]{Bhandari2020}, and Deep Synoptic Array \citep[DSA, $\pm 2.5 "$ accuracy;][]{Ravi2019}. This allows us to discover a large sample of bright FRB-like events with more accurate localization, including those located close to the Earth. In this section, we use parameters from BURSTT-2048 and CHIME to calculate a predicted number of gravitationally lensed FRB to be detected by the proposed BURSTT-2048. 

\subsection{Predicted number of FRBs to be detected by BURSTT}\label{sec:FRBs detected by BURSTT}
The specifications of BURSTT and CHIME relevant to the gravitational lensing search are summarised in Table \ref{tab:parameters}. Using the FoV and the SEFD of BURSTT and CHIME, we can estimate a value for the event rate of FRBs to be detected by BURSTT.
For simplicity, we assume the number of expected FRB numbers, $R$ is proportional to FoV and inversely proportional to the 3/2 power of SEFD as 
\begin{equation} \label{eq:09}
R \propto \frac{{\rm FoV}}{({\rm SEFD})^{3/2}}
\end{equation}
where the power-law index of the cumulative fluence distribution of CHIME FRBs. Observationally, the power is consistent with 3/2 within error according to the CHIME catalog paper \citep{CHIME2018} and it is also expected if there is no redshift evolution of FRB number density within Euclidean space. In our case, BURSTT looks at the nearby Universe. Therefore, such assumptions are likely the case. For CHIME, the FoV is $\sim$ 0.04 sr, and the SEFD = 100 Jy. For BURSTT-2048, FoV is $\sim$ 1 sr and the SEFD is $\sim$ 600 Jy.

Therefore, when we refer to 1,000 FRBs per year based on CHIME's predicted number \citep{CHIME2018}, we estimate that BURSTT-2048 will detect about 1,700 FRBs per annum, which outperforms CHIME in terms of FRB event rate. It should be also noted that BURSTT can localize all of these FRBs through Very-long-baseline interferometry (VLBI).

\begin{table*}
	\centering
	\caption{Specifications of BURSTT and CHIME relevant to the gravitational lensing search.}
	\label{tab:parameters}
	\begin{tabular}{ccccccc}
		\hline
		\hline
		Parameter & & Values&\\
		\hline
		\hline
		& BURSTT && CHIME&\\
		\hline
		Effective area & 40-200 m$^{2}$ (BURSTT-256) && 8,000 m$^{2}$ \\
		& 320-1600 m$^{2}$ (BURSTT-2048) &&&\\
		\hline
		& $\sim$ 1 sr &&  $\sim$ 0.04 sr \\
		FoV & $\sim100^{\circ}$ (E-W) && $2.5-1.3^{\circ}$ (E-W)\\
		& $\sim100^{\circ}$ (N-S) && $\sim100^{\circ}$ (N-S)\\
		\hline
		Bandwidth & 400 MHz && 400 MHz \\
		\hline
		Frequency range & 300-800 MHz && 400-800 MHz \\
		\hline
		SEFD & $\sim$ 5,000 Jy (BURSTT-256) && 100 Jy \\
		& $\sim$ 600 Jy (BURSTT-2048) &&&\\
		\hline
		Time resolution & $\sim$ 0.3 ms (adjustable to 1ns) && $\sim$ 2.56 $\mu$s (adjustable to 1 ns)\\
		\hline
		Localization accuracy & $\sim$ 1" &&  $\sim$ 2" \\
		& & & $\sim$ 25" (Pathfinder)\\
		\hline
		Polarization & Dual && Linearly dual \\
		\hline
  		E-W baseline & 800 km (Northern Taiwan to Hawaii) && 100 m \\
		\hline
  		N-S baseline & 300 km (Northern to Southern Taiwan) && 80 m \\
		\hline
	\end{tabular}
\end{table*}

\subsection{Predicted number of lensed FRBs to be detected by BURSTT}\label{sec:lensed FRBs detected by BURSTT}

\begin{figure}
\centering
	\includegraphics[width=\linewidth]{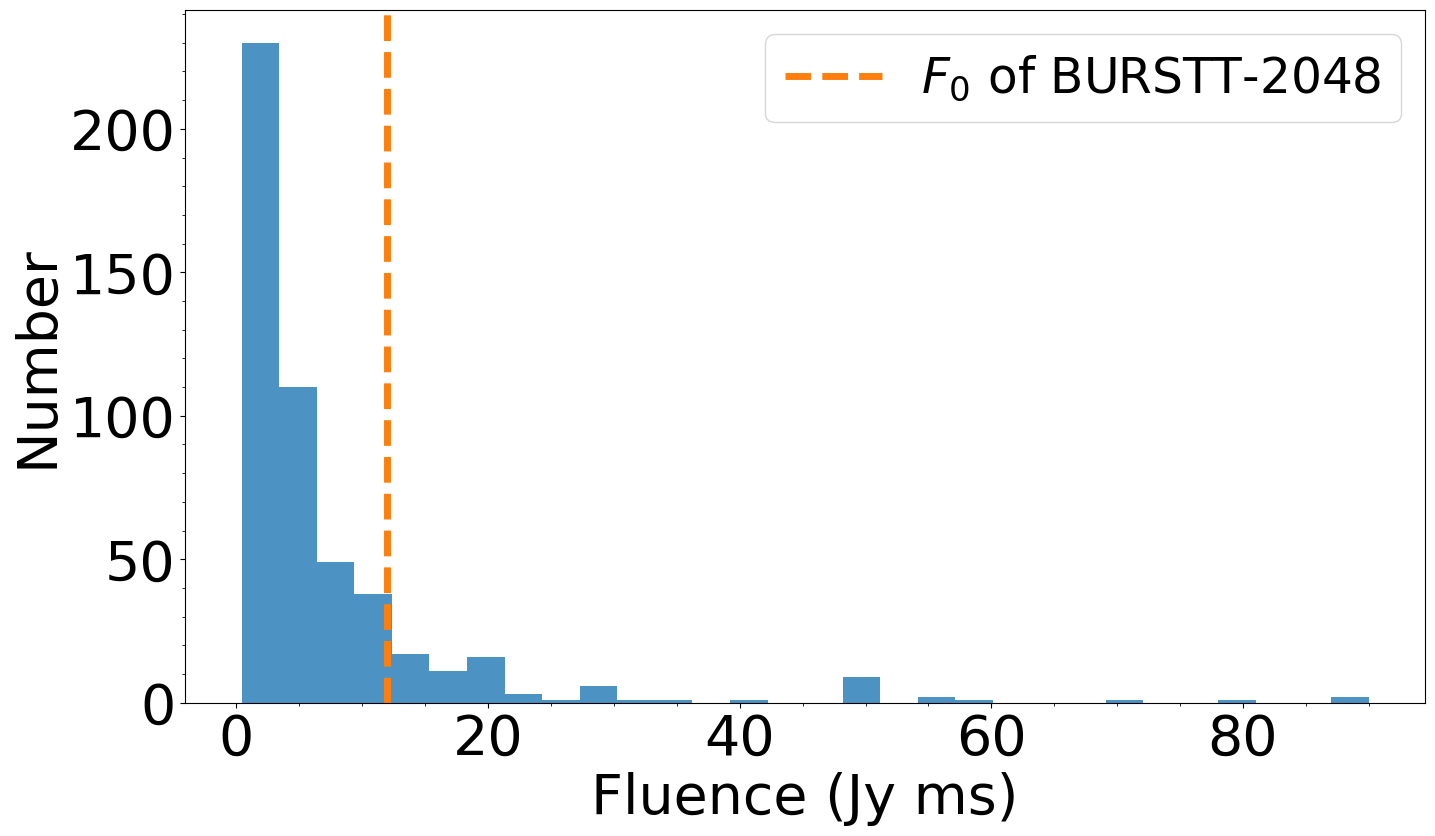}
    \includegraphics[width=\linewidth]{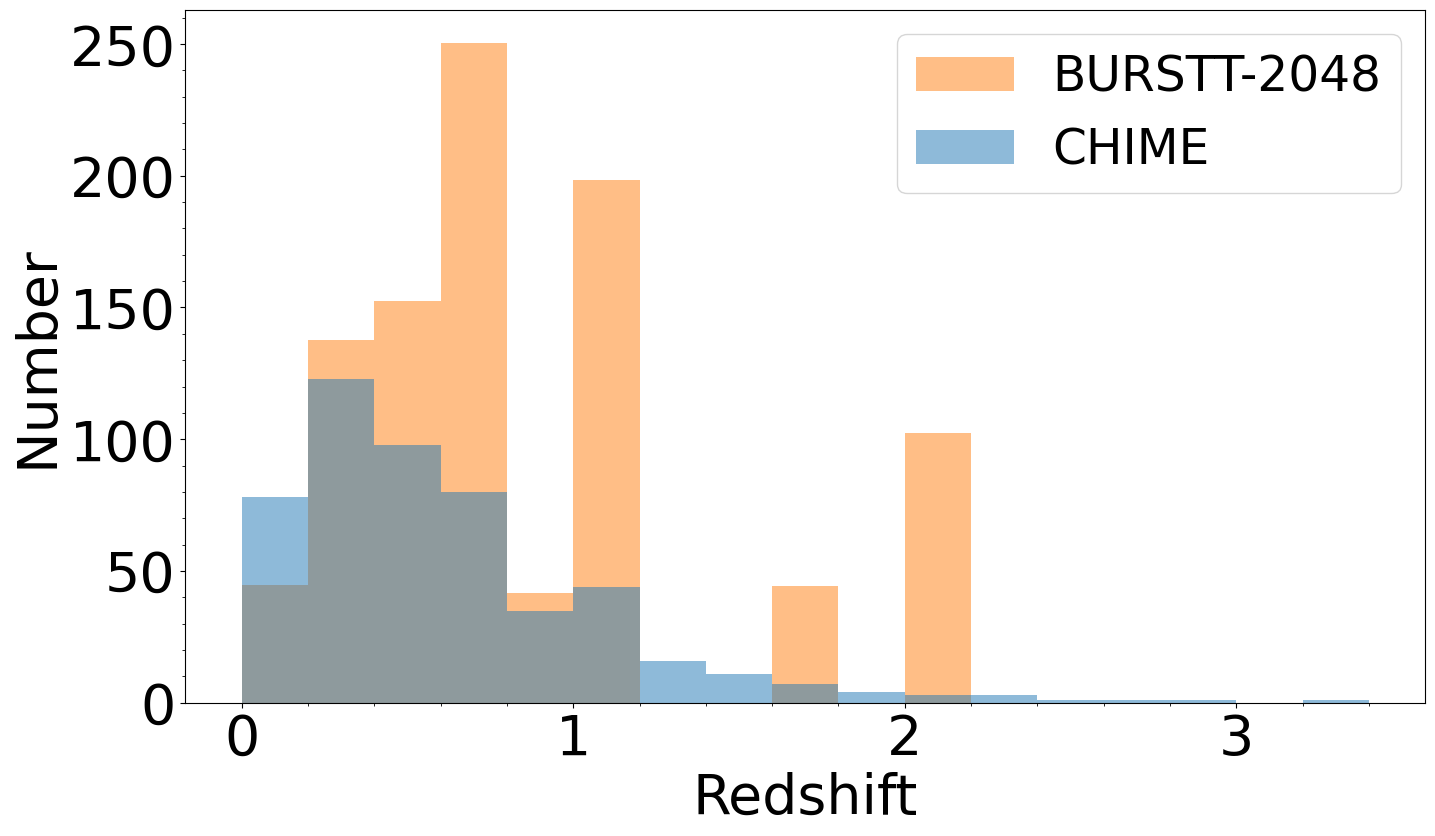}
    \caption{Fluence distribution of the CHIME FRB catalog \citep{CHIME2021} is shown in the upper panel. The nominal fluence detection threshold of BURSTT-2048 is marked with an orange dashed line in the upper panel. Redshift distribution of CHIME (blue), and BURSTT-2048 (orange) are shown in the lower panel.}
    \label{fig:fluencez_chime}
\end{figure}

Recall the Eq. \ref{eq:08}, given a redshift distribution function for FRBs, we can calculate the integrated optical depth due to a compact object of mass $M_{L}$, as
$$
\bar{\tau}\left(M_{L}\right)=\int d z \tau\left(z, M_{L}\right) N(z) .
$$
In our case, we obtained a redshift distribution from the latest CHIME catalog \citep{CHIME2021}. In this catalog, there are 536 FRBs. \citet{Hashimoto2022} estimated FRB's redshifts using their Dispersion Measure (DM). To estimate the fraction of FRB that can be detected by BURSTT, we need to estimate the nominal fluence detection threshold, $F_{0}$ of BURSTT. We use Eq. \ref{eq:10} from \citet{James2019}.
\begin{equation} \label{eq:10}
F_{0}=\frac{\sigma_{th}{\rm SEFD}}{\sqrt{2(t_{int})\Delta \nu}}
\end{equation}
where $\sigma_{th}$ represents the standard deviation of the detection threshold, $t_{int}$ represents the searching time resolution of the telescope, and $\Delta \nu$ represents the bandwidth of the telescope. We use $\sigma_{th} = 10$ and we estimate $F_{0}$ of BURSTT-2048 to be approximately 12 Jy ms. We put this threshold in the 536-FRB CHIME catalog and scale it with the FoV difference between CHIME and BURSTT(25 times better for BURSTT-2048). The estimated redshift distribution of BURSTT-detected FRBs is then recast. The fluence and redshift distributions of CHIME and BURSTT are shown in Fig. \ref{fig:fluencez_chime}. To calculate the integrated optical depth for each redshift bin, we calculate the optical depth $\tau$ for each FRB in the 536-FRB CHIME catalog. Recall Eq. \ref{eq:07}, $\tau$ is negative when $y_{\max } < y_{\min }^{2}$. This occurs when the lensing time scale, $\tau$ is small compared with the observational time resolution, $\bar{\Delta t}$. We cannot detect such signals and hence, there is no physical solution. In this case, the event is treated as a non-detection. After that, we sum the individual optical depth $\tau$ together in each redshift bin so we obtain the integrated optical depth in each redshift bin. These numbers are equivalent to the number of lensed FRBs in each redshift bin to be detected by BURSTT.

\begin{figure}
\centering
    \includegraphics[width=\linewidth]{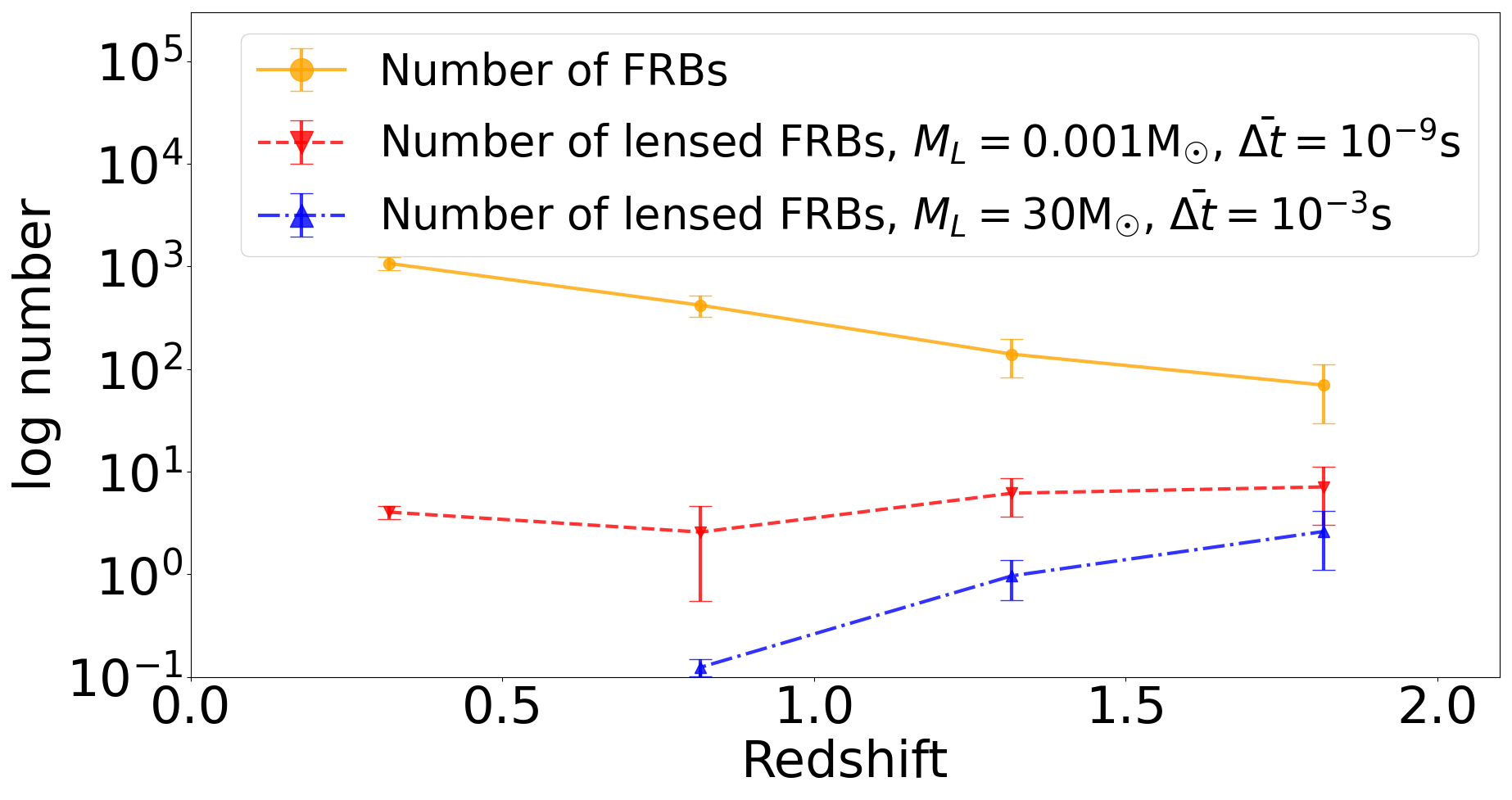}
    \caption{The number of lensed FRBs and the total number of FRBs for one-year operations of BURSTT-2048 in four redshift bins from $z=0$ to $z=2.2$. Each data point shows the bin center of the redshift bin. The orange line, green line, and red line show the redshift distribution of the total number of FRBs for one-year operations of BURSTT-2048, the redshift distribution of the number of lensed FRBs with $M_{L}$ = 0.001 $M_{\odot}$ and $\bar{\Delta t} = 1$ ns, and the redshift distribution of the number of lensed FRBs with $M_{L}$ = 30 $M_{\odot}$ and $\bar{\Delta t} = 1$ ms, respectively.}
    \label{fig:lensedfrb}
\end{figure}

We show BURSTT-2048's total number of FRBs and lensed FRBs per annum in four redshift bins from $z=0$ to $z=2.2$ in Fig. \ref{fig:lensedfrb}. In this plot, we show the number of detectable lensed FRBs with $M_{L}$=0.001$M_{\odot}$ and $\bar{\Delta t} = 1$ ns. which is the highest time resolution BURSTT-2048 can reach (H.-H. Lin 2022, private communication). In our estimation, we assume if all the dark matter is composed of PBHs ($f_{\mathrm{DM}}=1$), BURSTT-2048 can detect $\sim$ 24 lensed FRBs out of a total number $\sim$ 1,700 FRBs per annum. We also show the lensed number with $M_{L}$ = 30 $M_{\odot}$ and $\bar{\Delta t} = 1$ ms in blue. These are the parameters used in \citet{Munoz2016}. In this estimation, BURSTT-2048 can detect $\sim$ 5 lensed FRBs out of a total number of $\sim$ 1,700 FRBs per annum. Poisson errors propagated from the CHIME catalog \citep{CHIME2021} are included in the plot.

\subsection{The constraint of PBHs using lensed FRBs to be detected by BURSTT}\label{sec:constraint by BURSTT}
\begin{figure*}
\centering
	\includegraphics[width=\linewidth]{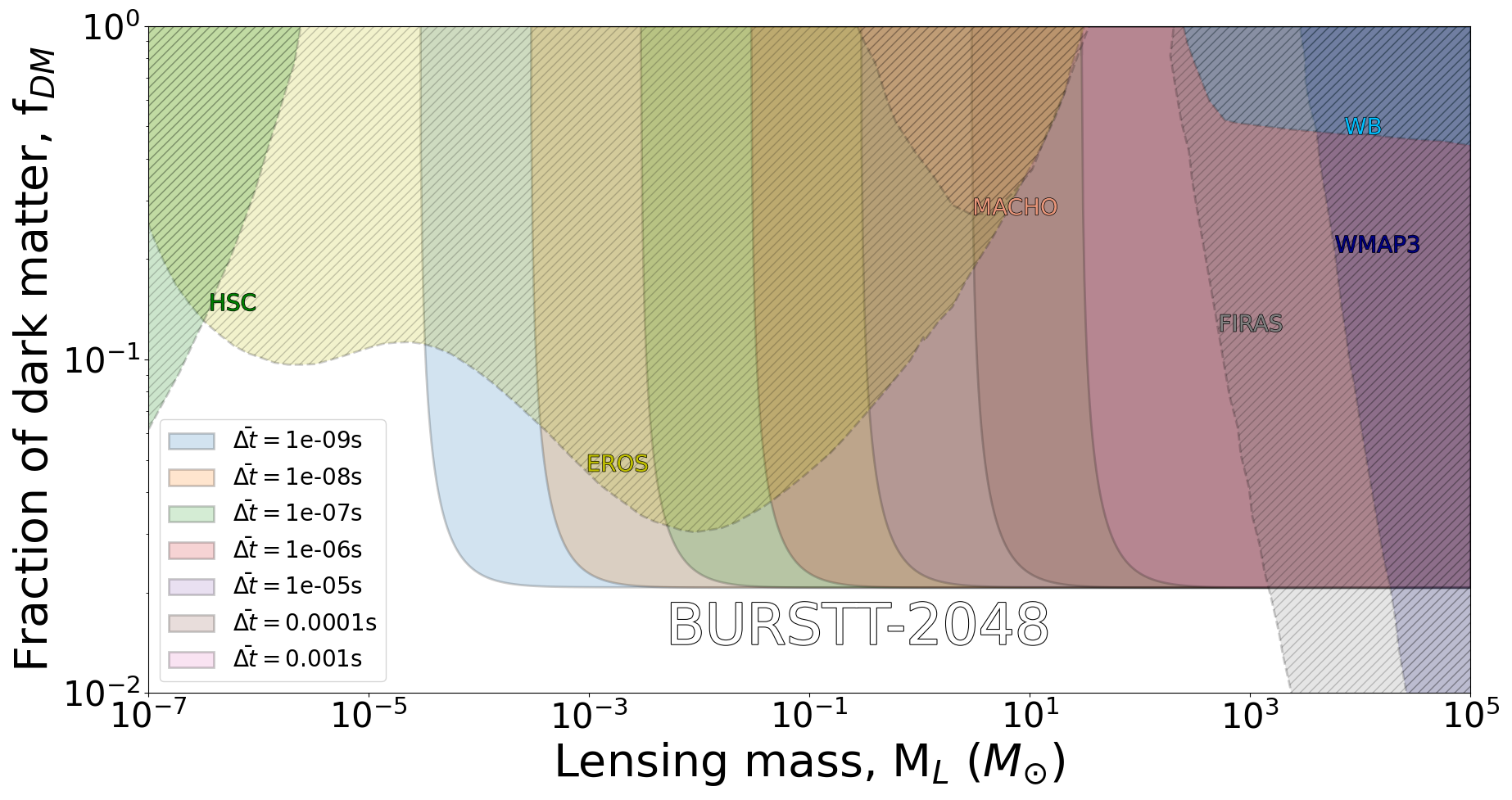}

    \caption{Fraction $f_{\mathrm{DM}}$ of dark matter in the form of point lenses of mass $M_{L}$, where the FRBs have a constant comoving density with a cutoff at $z_{\mathrm{cut}}=0.5$. Assuming a one-year operation, we show our constraints when we require a time delay $\bar{\Delta t}$ of, 1 ns, 10 ns, 0.1 $\mu$s, 1 $\mu$s, 10 $\mu$s, 0.1 ms, and 1 ms filled with} blue, orange, green, red, purple, brown, and pink, respectively. The condition of BURSTT-2048 is shown. We also show the current constraints from the MACHO Collaboration \citep{Alcock2000} filled with orange and hatches, with yellow and hatches the ones for the EROS Collaboration \citep{Tisserand2007}, with grey and hatches for the FIRAS \citep{Fixsen2002}, with green and hatches for the HSC \citep{Niikura2019}, with navy and hatches for the WMAP3 \citep{Ricotti2008} and with blue and hatches for the constraints from WB \citep{Quinn2010}.
    
    \label{fig:fdm}
\end{figure*}

In figure \ref{fig:fdm}, we show the regions in the parameter space of $f_{\mathrm{DM}}-M_{L}$ with a cutoff at $z_{\mathrm{cut}}=0.5$. Assuming the one-year operation of BURSTT-2048, we analyze the likelihood of at least one event for lensing time delays, $\bar{\Delta t}$ of, 1 ns, 10 ns, 0.1 $\mu$s, 1 $\mu$s, 10 $\mu$s, 0.1 ms, and 1 ms. In the same figure, we also show the current constraints to $f_{\mathrm{DM}}$ of the EROS Collaboration\citep{Tisserand2007}, MACHO Collaboration \citep{Alcock2000}, the COBE Far Infrared Absolute Spectrophotometer (FIRAS) data \citep{Fixsen2002}, the three-year Wilkinson Microwave Anisotropy Probe (WMAP3) \citep{Ricotti2008}, Subaru Hyper Suprime-Cam\citep[HSC;][]{Niikura2019}, and wide-binary (WB) disruption \citep{Quinn2010}. With BURSTT's ability to detect nanosecond bursts (H.-H. Lin 2022, private communication), we can constrain PBHs as dark matter down to $10^{-4}M_{\odot}$.

\section{Discussion}\label{sec:discussion}
\subsection{The number prediction of FRBs to be detected by BURSTT}
Referring to Section \ref{sec:lensed FRBs detected by BURSTT}, if we used the same parameters as \citet{Munoz2016}, BURSTT-2048 can detect 5 lensed FRBs (lensed by $30 \ M_{\odot}$ PBHs) out of 2200 FRBs with a time delay longer than 1 ms. The fraction of lensed FRBs is $\sim$ 0.2\% which is in the same order as \citet{Munoz2016}'s result (0.6\%).

\begin{figure*}
\centering
	\includegraphics[width=\linewidth]{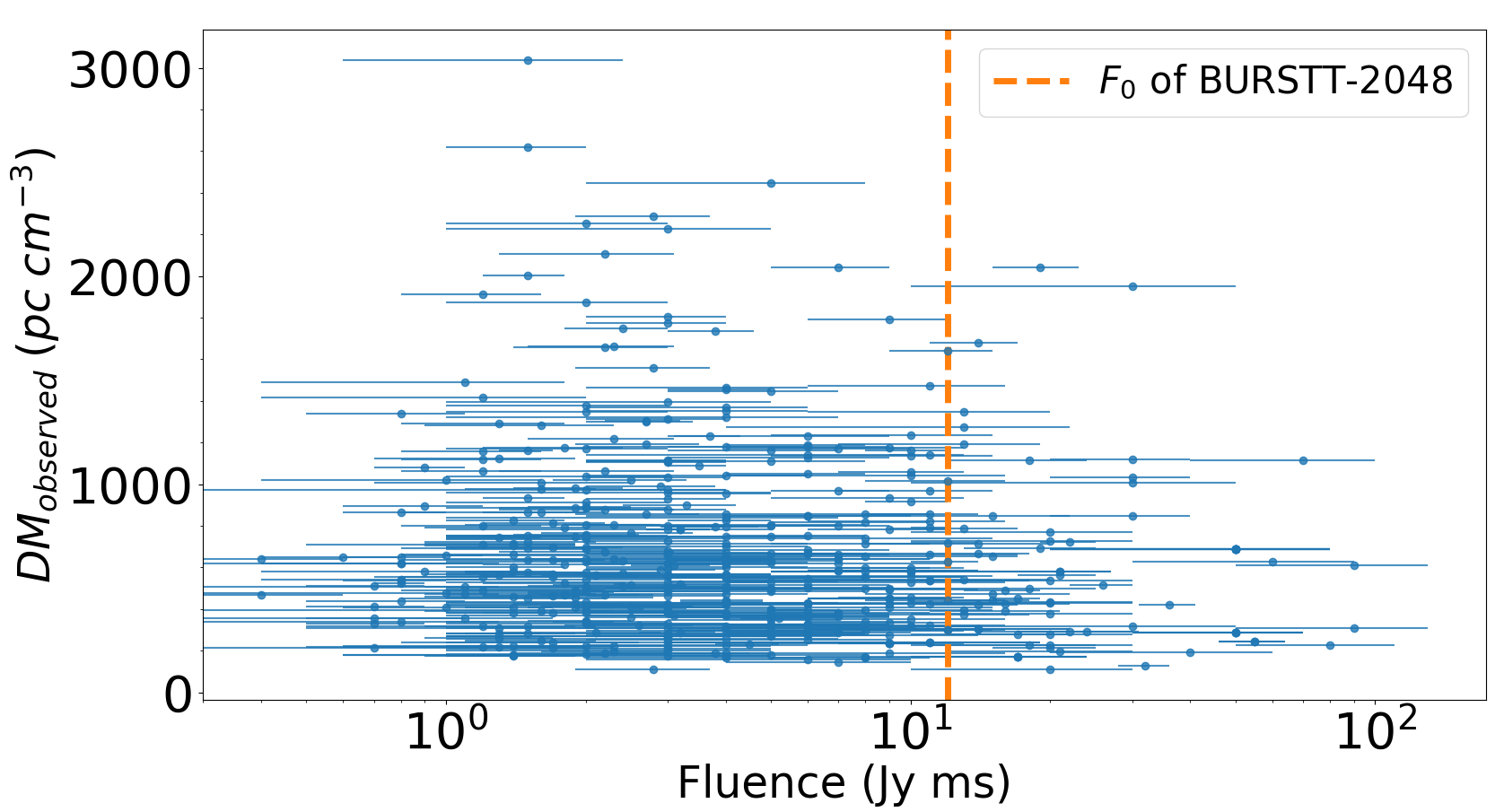}
    \caption{Observed DM versus fluence distribution for the CHIME catalog \citep{CHIME2021}. The nominal fluence detection threshold of BURSTT-2048, $F_{0}$, is marked with an orange dashed line. The x-axis is in logarithmic scale}
    \label{fig:dmf}
\end{figure*}

On the other hand, with the best time resolution (1ns) that BURSTT-2048 can reach, it can detect $\sim$ 24 lensed FRBs out of a total number $\sim$ 1,700 FRBs per annum if all the dark matter is in the form of $0.001M_{\odot}$ PBHs. This increases the fraction of lensed FRBs to $\sim$ 1.5\%. This fraction could be overestimated if we assume that BURSTT-2048 would detect FRBs at the very nearby Universe due to its sensitivity (SEFD $\sim$ 600 Jy).
 However, we should note that there are sources with high DM and high fluence in the CHIME FRB samples \citep{CHIME2021}. We show the distribution of observed DM versus fluence for the 536-FRB CHIME catalog \citep{CHIME2021} in Fig. \ref{fig:dmf}. 
In Fig. \ref{fig:dmf}, there are 35 FRBs with DM $>500$ pc cm$^{3}$ and fluence $>12$ Jy ms. These FRBs correspond to $\sim$ 48\% of FRBs with fluence $>12$ Jy ms. These are the extragalactic population of sources in the distant Universe ($z\gtrapprox 0.3-0.4$) that contribute to lensing optical depth in our estimation. As a result, the lensing fraction of 1.5\% would still be reasonable in our analysis based on the empirically derived redshift distribution of FRBs (Fig. \ref{fig:fluencez_chime}).

\subsection{The effect of plasma scattering}
\citet{Leung2022} discuss the effect of plasma scattering on the expected lensing event rate of FRBs. They constructed a two-screen model consisting of a gravitational lens and a plasma lens or scattering screen \citep{Leung2021}. Fig. 3 of \citet{Leung2022} shows the expected lensing event rate for 114 FRB events assuming that all dark matter is composed of PBHs with different lens masses. They show the expected lensing event rate changes for different effective distances of the plasma screen from the FRB source. The effect is maximal when the lens mass is in the range of about $10^{2}-10^{3}M_{\odot}$.

According to \citet{Leung2022}'s results, we can roughly calculate the decrease in the expected number of lensed FRBs for this paper. In the case of lensing mass $=0.001 M_{\odot}$, the expected number of lensed FRBs decreases by $\sim$ 5\%, $\sim$ 20\%, $\sim$ 60\%, and $\sim$ 90\% for 0.1 pc, 1 pc, 10 pc, and 100 pc of the plasma screen from the FRB source, respectively. This refers to the number of about 33 ($\sim$ 1.5\%), 27 ($\sim$ 1.2\%), 14 ($\sim$ 0.6\%), and 3 ($\sim$ 0.1\%) lensed FRBs (lensed fraction) per year, respectively.

\section{Conclusion}\label{sec:conclusion}
In this work, it has been demonstrated that coherent FRB lensing can constrain the constituents of the cosmological dark matter, e.g., PBHs with the parameters of BURSTT-2048. We consider compact objects which can be modeled by a point-mass lens \citep{Munoz2016}. Each FRB in the 536-FRB CHIME catalog is evaluated for optical depth $\tau$. To obtain the integrated optical depth in each redshift bin, we add the individual optical depth $\tau$ into each bin. Each of these numbers represents the number of lensed FRBs detected by BURSTT for each redshift bin. In our estimation, we assume if all the dark matter is composed of PBHs ($f_{\mathrm{DM}}=1$), BURSTT-2048 can detect up to $\sim$ 2 lensed FRBs out of a total number $\sim$ 1,700 FRBs per annum. With this amount of lensed FRBs and BURSTT's ability to detect nanosecond bursts, we can constrain PBHs as dark matter in the $10^{-4}-100M_{\odot}$ range. 


\begin{acknowledgments}

We are very grateful to the anonymous referee for many insightful comments.
TG and TH acknowledge the support of the National Science and Technology Council (NSTC) of Taiwan through grants 108-2628-M-007-004-MY3 and 110-2112-M-005-013-MY3/110-2112-M-007-034-, respectively. The BURSTT Project is funded by a grant from the NSTC (111-2123-M-001-008-).
We greatly appreciate Dr. Hsiu-Hsien Lin and Dr. Shotaro Yamasaki for providing useful comments.
\end{acknowledgments}

\bibliography{reference}{}
\bibliographystyle{aasjournal}


\end{CJK*}
\end{document}